\documentclass[12pt]{article}
\usepackage{latexsym}
\usepackage{graphicx}
\title{Boson-fermion unification, superstrings, and Bohmian mechanics}
\author{Hrvoje Nikoli\'c \\
Theoretical Physics Division, Rudjer Bo\v{s}kovi\'{c} Institute, \\
P.O.B. 180, HR-10002 Zagreb, Croatia \\
{\normalsize e-mail: hrvoje@thphys.irb.hr} \\
\makebox[1in]{} \\
}
\date{\today}
\begin{document}
\maketitle
\begin{abstract}
Bosonic and fermionic particle currents can be introduced in a 
more unified way, with the cost of introducing a preferred
spacetime foliation. Such a unified treatment 
of bosons and fermions naturally emerges from an 
analogous superstring current, showing that the preferred spacetime
foliation appears only at the level of effective field theory,
not at the fundamental superstring level.  
The existence of the preferred spacetime foliation
allows an objective definition of particles associated
with quantum field theory in curved spacetime.
Such an objective definition of particles makes 
the Bohmian interpretation of particle quantum mechanics
more appealing. The superstring current allows a 
consistent Bohmian interpretation of superstrings themselves,
including a Bohmian description of string creation and destruction 
in terms of string splitting.
The Bohmian equations of motion and the corresponding
probabilistic predictions are fully relativistic covariant
and do not depend on the preferred foliation.
\end{abstract}

\noindent
PACS: 03.65.Pm, 11.25.-w, 03.65.Ta \\
Keywords: Boson; fermion; unification; superstring; Bohmian mechanics

\maketitle

\section{Introduction}

All known elementary particles belong to one of the two fundamentally
different types: bosons and fermions. 
The unification of these two types of particles is one of the 
most fundamental issues in elementary-particle physics.
The best known theoretical principle for the unification of 
bosons and fermions is {\em supersymmetry} (see, e.g., \cite{SUSY}).
In particular, supersymmetric field theories are invariant 
with respect to a transformation that mixes bosonic and fermionic 
{\em fields}. Nevertheless, in field theories, each particle species
is described by a different field. Consequently, 
in supersymmetric field theories, 
bosonic and fermionic fields are still different fields.
Some supersymmetric field theories (typically, theories with only one supersymmetric
charge) can be formulated in terms of superfields that describe both
bosonic and fermionic fields as parts of a single superfield, but a superfield
formulation does not exist for all supersymmetric field theories. 

Different particle species are truly unified in {\em string} theories
\cite{zwie,GSW,polc}. In string theory, different particle species correspond 
to different states of a single object -- the string. Consequently, 
in supersymmetric string theory (shortly -- superstring theory), 
bosons and fermions
correspond to different states of a single string.

One of the differences between bosonic and fermionic particles 
is the fact that bosonic particles are described by 
{\em second}-order differential equations (e.g., Klein-Gordon equation), 
whereas fermionic particles are described by {\em first}-order
differential equations (e.g., Dirac equation). 
Consequently, the properties of the associated conserved 
particle current $j^{\mu}(x)$ 
are substantially different for bosons and fermions. The most notable
difference is the fact that $j^0(x)$ cannot be negative for 
fermions, whereas it can be negative (as well as positive) for bosons
(see, e.g., \cite{BD1}). 

In contrast to bosonic and fermionic particles, 
a superstring satisfies a set of equations called super-Virasoro constraints
that, in particular, contains stringy analogs of 
{\em both} the 
Klein-Gordon equation and the Dirac equation.   
In this paper we exploit this unifying feature of bosons and fermions
to introduce a new, more unified, theoretical framework to describe 
some aspects of bosons and fermions. More precisely, we introduce 
(i) particle currents for bosons and fermions in a more symmetric manner,
and (ii) a superstring current from which these bosonic and fermionic 
particle currents can be derived.

In construction of these particle and superstring currents we 
get confronted with two seemingly obscure features.
First, it turns out that the symmetric treatment of bosonic and 
fermionic particle currents requires an introduction of a 
preferred foliation of spacetime. It seems surprising
that a requirement of one symmetry (a symmetry between bosons 
and fermions) breaks another symmetry (a spacetime symmetry).
Is there any reasonable physical interpretation of such a 
preferred foliation of spacetime?
Second, the physical meaning of the particle and superstring 
currents themselves seems obscure too.     
What is the physical interpretation of these currents?
In this paper we argue that these seemingly obscure features 
acquire a natural interpretation in the context of 
{\em Bohmian} interpretation of particles and strings. 
To explain it, let us introduce the reader to the subject 
by shortly explaining what the significance 
of the Bohmian interpretation is, what the main problems
with this interpretation are, and how the results of this paper
help in solving them.

The Bohmian interpretation \cite{bohm,bohmPR1,bohmPR2,holrep,holbook} 
is the best known and most successful attempt to introduce a hidden-variable
completion of quantum mechanics (QM). According to this interpretation,
particle positions or fields have a continuous and deterministic dependence
on time even when they are not measured, while all quantum uncertainties
emerge from an ignorance of the actual initial conditions. This
interpretation offers a solution to the measurement problem of quantum
mechanics, but leads to several problems that make this interpretation
not widely accepted. 

First, owing to the Bell theorem \cite{bell}, any hidden-variable
completion of quantum mechanics must be nonlocal, which is usually
considered unacceptable in particle physics and field theory.
However, string theory contains 
certain nonlocal properties that have no analog in particle physics and 
field theory (see, e.g., \cite{nl1,nl2}), so, as observed in \cite{nikdual}, 
nonlocality does not seem to be an 
unacceptable feature from the point of view of string theory. 
Moreover, the nonlocal string Bohmian equation of motion seems to be a natural 
consequence of the world-sheet covariance of string theory
\cite{nikstring1}, which is an argument for the Bohmian interpretation
that has no analog in particle physics.   
    
Second, in the Bohmian interpretation of nonstring theories, 
it is not clear whether the fundamental objects are particles 
or fields. Owing to the fundamental difference between 
bosonic and fermionic particle currents discussed above,
as well as owing to the obscure nature of Grassmann-valued fields, 
it is often argued that bosons are fields, whereas fermions
are particles \cite{holrep,holbook}. Another boson-fermion asymmetric
Bohmian interpretation proposes that only bosonic fields have a 
true objective existence \cite{struyve}, so that all measurements
eventually reduce to a detection of bosons (e.g., photons). However, such 
boson-fermion asymmetric interpretations are clearly incompatible
with superstring theory. In this paper we formulate the 
Bohmian interpretation of superstrings, which in the low-energy 
limit necessarily leads to a boson-fermion symmetric Bohmian interpretation
of bosonic and fermionic particles. Thus, superstring theory supports
a picture according to which, at low energies, fundamental objects are 
bosonic and fermionic particles, rather than fields.

Perhaps the most undesirable feature of the Bohmian interpretation
is the fact that all attempts to reconcile nonlocality of the 
Bohmian interpretation with the requirement of relativistic covariance
lead, in one way or another, to a preferred foliation of spacetime 
(see, e.g., \cite{holrep,holbook,durrpra2,hort12,nikfpl1,nikfpl2,
nikddw12,niklosinj}). 
Typically, a need for a preferred foliation occurs because one needs to specify
a ``time coordinate'' with respect to which the nonlocal Bohmian influences
between spatially separated degrees of freedom are instantaneous.
On the other hand, most other interpretations of quantum mechanics
(which, of course, have their own problems not shared by the Bohmian
interpretation) do not need such a preferred foliation.
Therefore, the Bohmian interpretation would be much more appealing
if there were an interpretation-independent
theoretical argument for the existence of a preferred foliation,
or if it could be reformulated without need for a preferred foliation.
In a sense, in this paper we do both.
First, we argue that it is precisely the requirement 
of boson-fermion unification of the particle currents that 
provides such a missing argument for the existence of a preferred foliation,
which then makes the Bohmian interpretation more appealing as well.
Second, we formulate the Bohmian interpretation in such a manner
that it does not explicitly need a preferred foliation. Instead, the preferred
foliation is only needed to make an objective notion of particles 
consistent with quantum field theory (QFT) in curved spacetime, 
because otherwise particles are not well defined objects from the 
point of view of QFT in curved spacetime \cite{bd}.


The paper is organized as follows.
In the next section we study bosonic and fermionic particle currents
and use heuristic arguments (based on the assumption of  
boson-fermion unification) to construct a new fermionic particle 
current that leads to a preferred spacetime foliation in effective
field theory.
In Sec.~\ref{STRING} we introduce a natural superstring 
generalization of the particle current from which the unified 
bosonic and fermionic particle currents can be derived.
Sec.~\ref{BOHM} is devoted to the Bohmian interpretation, 
in which these particle and superstring currents attain a natural
interpretation.
In Sec.~\ref{PROB} we study the probabilistic interpretation
of nonrelativistic and relativistic wave functions and show that
it is consistent with the deterministic Bohmian interpretation.
The conclusions are drawn in Sec.~\ref{SECCONCL}.

In the paper, we use the spacetime signature $(+,-,\cdots ,-)$, 
while the units are chosen so that $\hbar=c=1$. 

\section{Particle currents for bosons and fermions}
\label{SEC2}

\subsection{The standard approach}
\label{SEC2.1}

A prototype of a wave equation for a bosonic particle is the 
Klein-Gordon equation
\begin{equation}\label{KG}
(\partial^{\mu}\partial_{\mu}+m^2)\varphi=0 .
\end{equation}
(Unless stated otherwise, in this section we assume that the number 
of spacetime dimensions is $D=4$.)
It describes particles with spin 0.
The associated conserved current is
\begin{equation}\label{KGcur}
j_{\mu}=i \varphi^* \!\stackrel{\leftrightarrow\;}{\partial_{\mu}}\! \varphi ,
\end{equation}
which is conserved in the sense that 
\begin{equation}\label{KGcur'}
\partial_{\mu}j^{\mu}=0 .
\end{equation}
We see that the Klein-Gordon equation is a second-order equation, 
and consequently, that the conserved current contains the first derivatives 
of $\varphi$. We also see that $j^0$ is not necessarily positive. Even if 
$\varphi$ is a superposition of plane waves with positive frequencies only,
$j^0(x)$ may still be negative locally. In particular, it implies that
$j^0(x)$ cannot be interpreted as a relativistic probability density.

A prototype of a wave equation for a fermionic particle is the
Dirac equation
\begin{equation}\label{dir}
(i\gamma^{\mu}\partial_{\mu}-m)\varphi=0 .
\end{equation}
It describes particles with spin $\frac{1}{2}$.
Here $\gamma^{\mu}$ are matrices that satisfy the Clifford algebra
\begin{equation}\label{clif}
\{ \gamma^{\mu},\gamma^{\nu} \}=2\eta^{\mu\nu} ,
\end{equation}
where $\eta^{\mu\nu}={\rm diag}(1,-1,-1,-1)$ is the Minkowski
metric. They also have the property
$\gamma^{0\dagger}=\gamma^{0}$, $\gamma^{i\dagger}=-\gamma^{i}$,
for $i=1,2,3$.
The wave function $\varphi$ is a 
4-component spinor.
The algebra (\ref{clif}) implies that $\varphi$ in (\ref{dir}) 
also satisfies the Klein-Gordon equation (\ref{KG}). On the other
hand, a solution of the Klein-Gordon equation (\ref{KG})
does not necessarily need to satisfy the Dirac equation (\ref{dir}).
Consequently, from solutions of (\ref{dir}) one can construct
a different conserved current that, in general, cannot be constructed
from solutions of (\ref{KG}). This current naturally 
associated with (\ref{dir}) is the Dirac current
\begin{equation}\label{dircur} 
j^{\mu}_{\rm D}=\bar{\varphi}\gamma^{\mu}\varphi ,
\end{equation}
where $\bar{\varphi}\equiv \varphi^{\dagger}\gamma^0$. It can be shown
(see, e.g., \cite{BD1}) that (\ref{dircur}) transforms as a vector,
and similarly, that $\bar{\varphi}\varphi$ transforms as a scalar.
We see that the Dirac equation (\ref{dir}) is a first-order equation,
and consequently, that the conserved current (\ref{dircur})
does not contain derivatives of $\varphi$.
Eq.~(\ref{clif}) implies $\gamma^0\gamma^0=1$, so (\ref{dircur})
implies $j^0_{\rm D}=\varphi^{\dagger}\varphi \geq 0$. 
In particular, it implies that
$j^0_{\rm D}(x)$ as given by (\ref{dircur}) can, potentially,
be interpreted as a relativistic probability density. 
(This, of course, does not prove that such an interpretation is 
physically correct.)

The moral of this subsection is that, in the standard approach, 
there is a large difference between bosonic and fermionic 
wave functions and associated conserved currents. In particular,
whereas the quantity $j^0_{\rm D}$ can, potentially, 
be interpreted as a probability density for the 
fermionic case, such an interpretation of $j^0$ does not work in the 
bosonic case. Of course, when $\varphi$ is interpreted as a 
bosonic or a fermionic {\em quantum field} (see, e.g., \cite{BD2}),
then such a probabilistic interpretation of $\varphi$ is no longer an issue.
However, as reviewed in \cite{nikmyth}, quantum field theory (QFT)
does not completely solve the problem, because it does not explain 
why the ``first-quantized" probabilistic interpretation of 
$\varphi^*(x)\varphi(x)$ is 
physically correct in the nonrelativistic limit. Moreover, 
from the string-theory perspective, it is possible that 
string field theory is {\em not} the correct approach to treat
the string interactions \cite{polcwhat}, 
which implies that, at low energies, 
particles may be more fundamental objects than fields.  
Thus, the goal of this paper is to introduce currents that treat 
bosons and fermions in a more symmetric way and to give 
an appropriate physical interpretation of such currents.  

\subsection{The Duffin-Kemmer approach}
\label{SECDuff}

To make boson and fermion currents more similar,
it has been proposed to replace the Klein-Gordon equation
for spin-0 particles with a first-order equation similar
to the Dirac equation. Such a first-order equation describing
massive spin-0 particles is known as the Duffin-Kemmer
equation \cite{kemmer}
\begin{equation}
(i\beta^{\mu}\partial_{\mu}-m)\Psi=0,
\end{equation}
where $\beta^{\mu}$ are matrices that satisfy
\begin{equation}
\beta^{\mu}\beta^{\nu}\beta^{\lambda} + 
\beta^{\lambda}\beta^{\nu}\beta^{\mu}=
\beta^{\mu}\eta^{\nu\lambda}+\beta^{\lambda}\eta^{\nu\mu}.
\end{equation}
The wave function $\Psi$ has 5 components that are related to the 
Klein-Gordon wave function $\varphi$ satisfying (\ref{KG}) via
\begin{equation}\label{PsiDK}
\Psi=\frac{1}{\sqrt{m}} \left( \begin{array}{c}
 \partial_{\mu} \varphi  \\
 m \varphi 
\end{array} \right) .
\end{equation}
The appropriate conserved current has been constructed in \cite{ghose}.
(For a review with an application to the Bohmian interpretation, see
also \cite{struy}.)
One first constructs the conserved energy-momentum tensor
\begin{equation}
\Theta^{\mu\nu}=m \bar{\Psi}(\beta^{\mu}\beta^{\nu}+\beta^{\nu}\beta^{\mu}
-\eta^{\mu\nu}) \Psi ,
\end{equation}
where $\bar{\Psi}\equiv \Psi^{\dagger} (2(\beta^0)^2-1)$.
The conserved current is then the Duffin-Kemmer current
\begin{equation}\label{curDK}
j^{\mu}_{\rm DK}=\Theta^{\mu\nu} n_{\nu} ,
\end{equation}
where $n^{\mu}$ is a unit future-oriented timelike vector.
In particular, by taking $n^{\mu}=(1,0,0,0)$, one finds
$j^0_{\rm DK}=\Theta^{00}=m\Psi^{\dagger}\Psi \geq 0$, which, potentially, 
may be interpreted as a probability density. However, a price to be paid 
is that it is necessary to introduce a special timelike direction
determined by the vector $n^{\mu}$. This direction 
defines a {\it preferred foliation of spacetime}.

We also note that bosonic and fermionic currents are still quite different,
as the fermionic current (\ref{dircur}) is not constructed from the 
energy-momentum tensor, whereas the bosonic current (\ref{curDK}) is.
In fact, as we shall see in Sec.~\ref{STRING}, the Duffin-Kemmer approach
reviewed above does not seem to be supported by superstring theory.
Therefore, in the next subsection, we propose a different approach. 
    
\subsection{A bosonlike approach}

In the preceding subsection, we have reviewed an approach that attempts 
to modify the bosonic current in a manner that makes it
more similar to the standard fermionic current. 
One of the consequences is a need to introduce a preferred
foliation of spacetime. 
In this subsection we propose a reversed strategy: we 
modify the fermionic current in a manner that makes it                  
more similar to the standard bosonic current.  
It will turn out that a need for an introduction 
of a preferred foliation of spacetime will emerge again.
However, such a unified treatment of bosonic and fermionic 
currents will turn out to be supported by superstring theory.

Consider a spin-$\frac{1}{2}$ wave function $\varphi$ that satisfies 
the Dirac equation (\ref{dir}). There are 4 ``natural" (but different) 
currentlike quantities that can be constructed from $\varphi$:
\begin{eqnarray}\label{4cur} 
& \bar{\varphi}\gamma_{\mu}\varphi , \;\;\;\;\;
\varphi^{\dagger} \gamma_{\mu}\varphi , & \nonumber \\
& i\bar{\varphi} \!\stackrel{\leftrightarrow\;}{\partial_{\mu}}\! \varphi,
\;\;\;\;\;
i\varphi^{\dagger} \!\stackrel{\leftrightarrow\;}{\partial_{\mu}}\! \varphi . &
\end{eqnarray}
Let us shortly discuss their properties. The first current
$\bar{\varphi}\gamma_{\mu}\varphi$ is the standard fermionic current
(\ref{dircur}). It is conserved, real, and transforms as a vector.
The second current $\varphi^{\dagger} \gamma_{\mu}\varphi$
is not conserved, the space components are not 
real, and it does not transform as a vector. The third current
$i\bar{\varphi} \!\stackrel{\leftrightarrow\;}{\partial_{\mu}}\! \varphi$,
just as the standard one, is conserved, real, and transforms as a vector.
The last current
$i\varphi^{\dagger} \!\stackrel{\leftrightarrow\;}{\partial_{\mu}}\! \varphi$
is conserved and real, but it does not transform as a vector.
So, which current is the correct one? Since our main guiding principle 
is the unification of bosons and fermions, the most natural
choice is the last one
\begin{equation}\label{curbos}
j_{\mu}=i\varphi^{\dagger} \!\stackrel{\leftrightarrow\;}{\partial_{\mu}}\!
\varphi ,
\end{equation}
because only this current has the same form as the bosonic current
(\ref{KGcur}). More precisely, (\ref{curbos}) looks just like a 
current of a many-component spin-0 wave. Therefore, we propose
that (\ref{curbos}) is the correct current for fermionic particles.
Indeed, as we shall see 
in Sec.~\ref{STRING}, such a unified particle current
for bosons and fermions naturally emerges from superstring theory.
Therefore, for the sake of an easier comparison with superstrings, we also 
write (\ref{curbos}) in a slightly different form. The spinor 
components $\varphi_m$ can be represented by 
$\varphi(m)=\sum_{m'} \delta(m-m')\varphi_{m'}$,
which allows to view the label $m$ as a continuous label
and, up to an inessential normalization factor,
to write (\ref{curbos}) as
\begin{equation}\label{curbosGr}
j_{\mu}(x)=i\int dm \, \varphi^*(x,m)
\!\stackrel{\leftrightarrow\;}{\partial_{\mu}}\!  
\varphi(x,m) .
\end{equation}
Eq.~(\ref{curbosGr}) will turn out to have a natural generalization
in superstring theory.

The problem with our choice (\ref{curbos}) is that 
this current does not transform as a
vector. Nevertheless, this problem can be avoided by writing it as
\begin{equation}\label{curuniv}
j_{\mu}  =  i\bar{\varphi}\gamma^0 
  \!\stackrel{\leftrightarrow\;}{\partial_{\mu}}\! \varphi
= i\bar{\varphi} (\gamma^{\nu}n_{\nu})
  \!\stackrel{\leftrightarrow\;}{\partial_{\mu}}\! \varphi ,
\end{equation}
where $n^{\mu}=(1,0,0,0)$.
From the last expression in (\ref{curuniv}), 
we see that this current transforms as a vector.
However, the price paid is that it is necessary to introduce 
a preferred foliation of spacetime induced by the unit future-oriented 
timelike vector $n^{\mu}$. 
The physical meaning of such a preferred foliation
is discussed in the next subsection.

\subsection{Preferred foliation and particles vs fields}
\label{SECpfpf}

It is interesting to observe that both attempts (the one introduced
in the preceding subsection as well as the one reviewed in 
Sec.~\ref{SECDuff})
to introduce particle currents for bosons and fermions in a more 
similar way led to a preferred foliation of spacetime.
Does this fact bring a deeper physical message? In this subsection we 
propose an answer to this question and support it by
some quantitative and qualitative arguments. 

In Eq.~(\ref{curuniv}), the need for an introduction of a preferred
foliation of spacetime emerged from the observation that otherwise
the current (\ref{curbos}) does not transform as a vector.
But why exactly (\ref{curbos}) does not transform as a vector?
This is because $\varphi$ is a spinor (not merely a collection
of scalars) that transforms nontrivially
under Lorentz transformations. But why must $\varphi$ transform in
such a nontrivial way? This transformation is derived \cite{BD1}
from the requirement that the Dirac equation (\ref{dir})
should be covariant. However, the crucial {\em assumption} in this
derivation is that $\gamma^{\mu}$ is {\em not} a vector, but a collection of 
fixed matrices. 

Conversely, if $\gamma^{\mu}$ were assumed to be a vector, then
the covariance of (\ref{dir}) would be
consistent with the possibility that $\varphi$ does 
{\em not} transform as a spinor, but merely as a collection of 
Lorentz scalars. 
(This should not be confused with a more common approach based on 
vielbeins \cite{GSW,bd}, in which $\gamma^{\mu}$ transforms as 
a vector under general coordinate transformations.
In this picture, $\varphi$ transforms as a spinor under {\em internal}
Lorentz transformations, which does not comply with our unifying
picture of bosons and fermions. The relation between these two
pictures will become clearer in Sec.~\ref{STRING}.)
In this case, the current (\ref{curbos}) 
would be a genuine vector without the need for an introduction of a preferred
foliation of spacetime. Could such a reinterpretation of the current
(\ref{curbos}) be the correct one?  
As we shall see in Sec.~\ref{STRING}, in superstring theory
(which, indeed, provides an explicit unification of bosons and fermions)
$\gamma^{\mu}$ really emerges from a vector quantity and 
$\varphi$ really emerges from a quantity that transforms as a Lorentz scalar.
Thus, {\em from the fundamental superstring point of view,
the current (\ref{curbos}) is naturally interpreted as a genuine vector}.   

Now, with such a reinterpretation of $\gamma^{\mu}$ and $\varphi$,
it is the Dirac current (\ref{dircur}) 
\begin{equation}\label{curD2}
j^{\mu}_{\rm D}=\varphi^{\dagger}\gamma^0\gamma^{\mu}\varphi
\end{equation}
that does not transform as a vector, unless a preferred foliation with the
normal $n^{\mu}$ is used, so that we can write 
\begin{equation}\label{curD3}
j^{\mu}_{\rm D}=\varphi^{\dagger}(\gamma^{\nu}n_{\nu})\gamma^{\mu}\varphi .
\end{equation}
Does the fact that now (\ref{curD2}) does not transform as a vector
represent a physical problem? For free particles, there are no substantial
problems. For example, the Dirac equation can be derived from a 
scalar Lagrangian density 
$\varphi^{\dagger}(i\gamma^{\mu}\partial_{\mu}-m)\varphi$, while 
the Dirac current (\ref{curD2}) does not play a particular role
for free particles. However, the Dirac current plays an important role 
for particle {\em interactions}. For example, 
the coupling between a charged fermion and the electromagnetic field 
$A_{\mu}$ is given by the interaction Lagrangian proportional to 
$A_{\mu}j^{\mu}_{\rm D}$, which now is not a scalar, unless a preferred 
foliation is introduced. This means that a true problem occurs only when
one attempts to describe particle {\em interactions}. However, 
from the fundamental superstring point of view, this does not represent a 
problem at all because, in perturbative superstring theory, all 
interactions are described by the {\em free} superstring Lagrangian
\cite{GSW,polc}.
Thus, the problem occurs only at the level of low-energy {\em effective}
QFT description of interactions.

The observations above also lead to an interesting reinterpretation
of the theory of quantum particles and fields in classical 
curved backgrounds. In the usual interpretation of 
QFT in curved spacetime \cite{bd}, the fundamental quantity is assumed to be 
a quantum field operator $\hat{\phi}(x)$, while the concept of a particle 
is regarded as emergent. It turns out that the notion of particles 
depends on the choice of the time coordinate with respect to which 
the notion of positive and negative frequencies is defined. In other
words, in general, the notion of a particle is not well defined.
An alternative is to introduce a preferred foliation of spacetime, 
which then allows to introduce a unique notion of particles described
by local covariant field operators of particle currents \cite{nikcur}.  
The point is that, when the field is assumed to be fundamental, then
it is the particle (not the field) that is problematic and requires 
a preferred foliation. On the other hand, if one assumes that the 
fundamental quantity is not a field but a superstring, then the low-energy 
object described by the fundamental quantity is a particle, not a field.
(Of course, if the fundamental quantity at high energies is not a 
string but a string field, then the conclusion above changes.
Here, however, we consider the possibility that string field theory
is {\em not} the correct approach to treat strings at a nonperturbative level
\cite{polcwhat}.)   
In other words, from the point of view of the fundamental perturbative
superstring theory, at low energies particles are more fundamental objects
than fields, while fields only serve as auxiliary mathematical objects 
useful for a description of particle interactions at low energies.
This means that in curved spacetime particles should be well defined
even without a preferred foliation, while it is the concept of 
an associated quantum field that is not well defined unless a 
preferred foliation is introduced. Indeed, such a reinterpretation
of QFT in curved spacetime is consistent with our conclusion above 
that a need for a preferred foliation in the Dirac current (\ref{curD3})
occurs only at the level of low-energy {\em effective}
QFT description of interactions.  

To further clarify the conceptual difference between quantum particles 
and fields, consider a free hermitian quantum field operator
$\hat{\phi}(x)$ satisfying the Klein-Gordon equation (\ref{KG}).
To attribute a particle interpretation to the field $\hat{\phi}(x)$,
one is forced to expand it as
\begin{equation}\label{eqf}
\hat{\phi}(x)=\sum_k \hat{a}_k\varphi_k(x) + \hat{a}_k^{\dagger}\varphi_k^*(x) ,
\end{equation}
where the functions $\varphi_k(x)$ and $\varphi_k^*(x)$ 
have positive and negative norms, respectively, and constitute
some complete orthonormal set of solutions to (\ref{KG}).
In the particle interpretation,
the operators $\hat{a}_k$ and $\hat{a}_k^{\dagger}$ 
are interpreted as the destruction and creation 
operators, respectively. 
However, in a curved-spacetime generalization of (\ref{KG}),
the choice of the basis $\{ \varphi_k,\varphi_k^* \}$
is not unique, which implies that the definition of particles
based on the destruction and creation operators $\hat{a}_k$ and
$\hat{a}_k^{\dagger}$ is also not unique. In particular, if 
$|\Phi\rangle$ is a QFT state, then the positive-norm 1-particle wave function
is \cite{ryder}
\begin{equation}\label{wffield}
\varphi(x)=\langle 0|\hat{\phi}(x)|\Phi\rangle ,
\end{equation}
where $|0\rangle$ is the vacuum defined by $\hat{a}_k|0\rangle=0$.
(For a generalization to the many-particle wave function and spin 
$\frac{1}{2}$,
see also \cite{schw,nikfpl1,nikfpl2}.)
Thus, the 1-particle wave function $\varphi(x)$ is also not unique.
In the conventional interpretation, the field $\hat{\phi}(x)$ is,
by assumption, always well defined, so it is the particle
wave function $\varphi(x)$ on the left-hand side of (\ref{wffield})
that cannot be defined without a preferred foliation of spacetime.
However, in our reinterpretation, fields and wave functions exchange their
roles. Now the particle wave function
$\varphi(x)$ is assumed to be fundamental and well defined.
The field then only plays an auxiliary role in writing the wave function
in terms of fields as in (\ref{wffield}). To do this, one is forced
to expand the positive-norm 1-particle wave function $\varphi(x)$ as
\begin{equation}\label{eqp}
\varphi(x)=\sum_k c_k\varphi_k(x) ,
\end{equation}
where $c_k$ are some complex coefficients, so that one can {\em define}
the auxiliary field operator through the expansion
(\ref{eqf}) that allows the identification
(\ref{wffield}). Now it is the field operator (\ref{eqf}) that depends
on the choice of the basis $\{ \varphi_k,\varphi_k^* \}$, while
the expansion (\ref{eqp}) of the wave function is considered purely
conventional. (We shall discuss it more systematically elsewhere.)   
Although we are not able to prove explicitly that the preferred foliation
needed for the unification of bosonic and fermionic currents must
necessarily coincide with the preferred foliation needed
for the relation between particles and fields, the simplest
possibility is, indeed, that these two foliations coincide.

To summarize, we conjecture that the preferred foliation that 
emerges from our attempt to introduce a unifying current for 
bosons and fermions is a manifestation of the fact that a 
QFT description of particle interactions is not fundamental.
The implications on the theory of particles and fields in 
curved spacetime will be studied in detail elsewhere.
The fundamental physical origin of this preferred foliation
is still unclear, but the sections that follow provide a further insight.
In the next section we show how the unifying current appears more 
naturally in superstring theory, while
in Sec.~\ref{BOHM} we propose an interpretation of the particle and
superstring currents.

\section{Superstring current}
\label{STRING}

\subsection{Elements of superstring theory}
 
To make this paper accessible to readers not familiar with 
superstring theory, following \cite{GSW}, 
in this subsection we briefly outline 
some basics of superstring theory. 
The shape of a classical string evolving 
in spacetime is described by real functions $X^{\mu}(\sigma,\tau)$, 
$\mu=0,1,\ldots, D-1$, where $\sigma\in [0,\pi]$ is the parameter
along the string, and $\tau$ parametrizes time. 
A superstring is also equipped by 
Grassmann-valued coordinates $\psi^{\mu}(\sigma,\tau)$, where each
$\psi^{\mu}$ is a 2-component world-sheet Majorana spinor 
\begin{equation}
\psi^{\mu}=\left(
\begin{array}{c}
\psi_1^{\mu} \\
\psi_2^{\mu}
\end{array} \right)
\equiv
\left(
\begin{array}{c}
\psi_-^{\mu} \\
\psi_+^{\mu}
\end{array} \right) . 
\end{equation}
In addition, $\psi^{\mu}$ transforms as a vector with respect to
Lorentz transformations in spacetime. 
(Our convention for various types of indices is summarized in 
Table 1.)
Introducing the notation $\tau=\sigma^0$, $\sigma=\sigma^1$, 
the action of a superstring can be written as 
\begin{equation}\label{Act}
A=-\frac{1}{2}\int d^2\sigma 
[(\partial_{\alpha} X^{\mu})(\partial^{\alpha} X_{\mu})
+i\bar{\psi}^{\mu}\rho^{\alpha}\partial_{\alpha}\psi_{\mu}] ,
\end{equation}
where $\alpha=0,1$, $\rho^{\alpha}$ are $2\times 2$ matrices
\begin{equation}
\rho^0=\left( 
\begin{array}{cc}
0 & -i \\
i & 0
\end{array} \right) , \;\;\;\;
\rho^1=\left( 
\begin{array}{cc}
0 & i \\
i & 0
\end{array} \right) ,
\end{equation}
satisfying the 2-dimensional Clifford algebra
$\{ \rho^{\alpha}, \rho^{\beta} \}=2\eta^{\alpha\beta}$, 
and $\bar{\psi}\equiv \psi^{\dagger}\rho^0$.
This action is invariant under the global 
world-sheet supersymmetry transformations
\begin{eqnarray}\label{transf}
& \delta \psi^{\mu} = -i (\partial_{\alpha} X^{\mu})\rho^{\alpha}\epsilon ,
\;\;\;\;
\delta \bar{\psi}^{\mu} = i (\partial_{\alpha} X^{\mu})
\bar{\epsilon}\rho^{\alpha} , & \nonumber \\
& \delta X^{\mu} = \bar{\epsilon}\psi^{\mu} , &
\end{eqnarray}
where $\epsilon$ is a constant infinitesimal Grassmann-valued 2-component 
Majorana spinor. 

\begin{table}[t] 
\label{table1}
\[
\begin{tabular}{|c|c|c|} 
\hline
manifold      & $\;$vector indices$\;$   & $\;$spinor indices$\;$ \\ \hline
spacetime           & $\mu$, $\nu$             & $m$            \\ 
$\;$world-sheet$\;$  & $\alpha$, $\beta$    & $b$            \\ \hline
\end{tabular}
\]
\caption{Summary of the indices conventions.}
\end{table}

When $\epsilon$ is not constant, then (\ref{Act}) is no longer invariant
under (\ref{transf}). Instead, the variation of the action $\delta A$ 
is proportional to 
$\int d^2\sigma \, (\partial_{\alpha}\bar{\epsilon}) S^{\alpha}$, 
where
\begin{equation}\label{superc}
S_{\alpha}=\frac{1}{2}\rho^{\beta}\rho_{\alpha}\psi^{\mu}
\partial_{\beta}X_{\mu} 
\end{equation}
is the supercurrent. (In the literature, $S_{\alpha}$ is usually 
denoted by $J_{\alpha}$, but we change the notation because
in this paper we will introduce a new superstring current 
$J_{\mu}$ that generalizes the particle current $j_{\mu}$ of 
Sec.~\ref{SEC2}.)
In superstring theory, one actually requires local (not only global)
world-sheet supersymmetry. This means that $S_{\alpha}$ should vanish
for physical states. Similarly, the requirement that the action should be 
invariant under reparametrizations of the world-sheet coordinates 
$\sigma^{\alpha}$ implies that the world-sheet energy-momentum tensor
$T_{\alpha\beta}$ should also vanish for physical states.
(To understand this, note that
in arbitrary world-sheet coordinates the Minkowski metric $\eta_{\alpha\beta}$
generalizes to a more general metric $h_{\alpha\beta}(\sigma^0,\sigma^1)$, 
while the measure $d^2\sigma$ generalizes to $d^2\sigma \sqrt{|h|}$, 
where $h$ is the determinant of  $h_{\alpha\beta}$. Since the action (\ref{Act})
generalized in this way must be reparametrization invariant, we must have
$\delta A/\delta h_{\alpha\beta}=0$. This implies that the energy-momentum tensor
defined as $T^{\alpha\beta}=\left(-2/\sqrt{|h|}\right)\, \delta A/\delta h_{\alpha\beta}$
must vanish.) 
It is convenient to introduce the light-cone world-sheet coordinates
$\sigma^{\pm}=(\tau \pm \sigma)$, so that the constraints
on $S_{\alpha}$ and $T_{\alpha\beta}$ can be written as
\begin{equation}
T_{++}=T_{--}=0 , \;\;\;\;
S_+=S_-=0 .
\end{equation}
From these local constraints one can construct an infinite set of
global constraints. For open strings, these global constraints are
\begin{equation}\label{constrLF}
L_n=0 , \;\;\;\; F_l=0 ,
\end{equation}
where
\begin{equation}\label{gen1}
L_n\equiv \int_0^{\pi} d\sigma (e^{in\sigma} T_{++}
+ e^{-in\sigma} T_{--}) ,
\end{equation}
\begin{equation}\label{gen2}
F_l\equiv \sqrt{2} \int_0^{\pi} d\sigma (e^{il\sigma} S_{+}
+ e^{-il\sigma} S_{-}) .
\end{equation} 
Here $n$ are integers, while $l$ are integers or half-integers, 
corresponding to $\psi^{\mu}(\sigma)$ that satisfies the 
Ramond boundary condition 
$\psi^{\mu}_+(\pi)=\psi^{\mu}_-(\pi)$ (R sector) or the
Neveu-Schwarz boundary condition 
$\psi^{\mu}_+(\pi)=-\psi^{\mu}_-(\pi)$ (NS sector), respectively.
The global quantities $L_n$ and $F_l$ are referred to as 
super-Virasoro generators.
To be more precise, these super-Virasoro generators describe open strings.
In the case of closed strings, there are two copies of 
super-Virasoro generators above, corresponding to right-moving
and left-moving strings. Thus, for closed strings we have 
R-R, NS-NS, R-NS, and NS-R sectors. 

In the quantum theory, the superstring coordinates become the operators
$\hat{X}^{\mu}$ and $\hat{\psi}^{\mu}$. They satisfy the canonical
equal-time (anti)commutation relations
\begin{equation}\label{comrel1}
[\hat{X}^{\mu}(\sigma),\hat{P}^{\nu}(\sigma')]=
-i\eta^{\mu\nu}\delta(\sigma-\sigma') ,
\end{equation}
\begin{equation}\label{comrel2}
\{ \hat{\psi}_b^{\mu}(\sigma),\hat{\psi}_{b'}^{\nu}(\sigma') \}=
-\eta^{\mu\nu}\delta_{bb'}\delta(\sigma-\sigma') ,
\end{equation}  
where $b=1,2$ is the world-sheet spinor index and 
$\hat{P}^{\mu}=\partial \hat{X}^{\mu}/\partial\tau$. 
Thus, the super-Virasoro generators (\ref{gen1}) and
(\ref{gen2}) also become the operators, while the constraints  
(\ref{constrLF}) become the constraints on physical states
\begin{eqnarray}\label{constrLFop}
& \hat{F}_l |\Psi\rangle=0 , \;\; l\geq 0 , & \nonumber \\
& \hat{L}_n |\Psi\rangle=0 , \;\; n>0 , & \nonumber \\
& (\hat{L}_0-a) |\Psi\rangle=0 . & 
\end{eqnarray}
Here $a$ is a constant that depends on the ordering of the operators.
When the normal ordering is used, then $a=1/2$ for the NS sector 
and $a=0$ for the R sector. It turns out that the 
quantum super-Virasoro constraints (\ref{constrLFop}) are consistent 
only for $D=10$. 
The NS sector corresponds to bosonic states in spacetime, while the R sector 
corresponds to fermionic states in spacetime. However, not all states satisfying
(\ref{constrLFop}) are physical. The physical spectrum is a truncated
one obtained from the spectrum above by the so-called GSO projection, 
which introduces further symmetry of the spectrum -- 
spacetime supersymmetry \cite{GSW,polc}. 

The correspondence between strings and particles can be seen
by considering the modes with $n=l=0$.
From (\ref{gen1}) we see that $L_0$ is the Hamiltonian 
$\int d\sigma\, T_{00}$ 
with a term quadratic in the momenta $P^{\mu}$
(see also (\ref{Hamilt}) below),
so the last equation in (\ref{constrLFop}) 
represents a stringy analog of the Klein-Gordon equation.
In addition, the operator $\hat{F}_0$ satisfies
$\hat{F}_0\hat{F}_0=\hat{L}_0$, so the equation $\hat{F}_0 |\Psi\rangle=0$
of the fermionic R sector can be viewed as a stringy analog 
of the Dirac equation. 
Note, however, that a stringy analog of the
Klein-Gordon equation is satisfied by both sectors, 
while that of the Dirac equation is satisfied only by one of the 
sectors. This fact will be crucial in construction of the unifying
superstring current in the next subsection.

\subsection{Schr\"odinger picture and the superstring current}

In this subsection, we want to construct a superstring current 
that represents an analog of the particle current $j_{\mu}$ 
discussed in Sec.~\ref{SEC2}. A natural starting point is to 
write the last equation in (\ref{constrLFop}) in a form 
that resembles the Klein-Gordon equation more explicitly, 
and similarly for the equation $\hat{F}_0 |\Psi\rangle=0$
that resembles the Dirac equation. In particular, the 
momentum $\hat{P}_{\mu}$ should be represented by a derivative operator, 
analogously to the particle momentum operator 
$\hat{p}_{\mu}=i\partial_{\mu}$. For that purpose, we need 
to introduce the Schr\"odinger picture of quantum superstring theory.

Before starting with the explicit construction, the following observations 
are crucial. In analogy with the standard approach to particles
in Sec.~\ref{SEC2.1}, 
one could attempt to construct one superstring current for the 
bosonic NS sector and another superstring current for the 
fermionic R sector. However, according to the quantum superposition
principle, a general quantum state is a {\em superposition of 
states from both sectors}. This fact does not have an analog 
at the level of (first-quantized) particle wave functions; 
a superposition of solutions of the Klein-Gordon and the 
Dirac equation does not make sense. This reflects the fact 
that in particle physics, as well as in quantum field theory,
different particle species correspond to genuinely different objects. 
On the other hand, in superstring theory 
different particle species are only different states of the 
same object -- the superstring. Therefore, in a consistent 
superstring theory {\em there should be only one superstring current}
describing both NS and R sectors, as well as their arbitrary superpositions.
Consequently, as both sectors satisfy the
stringy analog of the Klein-Gordon equation,
it is the stringy analog of the Klein-Gordon equation,
and not that of the Dirac equation, from which the superstring current 
should be constructed. 

Now we are ready for the explicit construction.
From the action (\ref{Act}) one finds the Hamiltonian
\begin{equation}\label{Hamilt}
H=-\frac{1}{2} \int d\sigma [P^{\mu}P_{\mu} 
+(\partial_{\sigma}X^{\mu}) (\partial_{\sigma}X_{\mu})
-i\psi^{{\sf T}\mu}\rho^0\rho^1\partial_{\sigma}\psi_{\mu} ], 
\end{equation}
where $P^{\mu}$ and $\psi^{\mu}$ are functions of $\sigma$,
$\partial_{\sigma}\equiv \partial/\partial\sigma$, and 
$\psi^{\dagger}=\psi^{{\sf T}}$ for a Majorana spinor.
In the Schr\"odinger picture, the operator $\hat{P}_{\mu}(\sigma)$
is represented by the functional derivative
\begin{equation}\label{eqn34}
\hat{P}_{\mu}(\sigma)=i\frac{\delta}{\delta X^{\mu}(\sigma)} ,
\end{equation}
which is consistent with (\ref{comrel1}).
Various boundary conditions are imposed
on the states $\Psi$, not on the operators. 
The explicit representation of the operator $\hat{\psi}^{\mu}(\sigma)$
will not be needed here, but we note that it is a 
direct sum of the corresponding operators for all sectors
(such as open R sector, closed R-R sector, etc.).
For our purposes, it is sufficient to know that $\hat{\psi}^{\mu}$ acts
in some Hilbert space with indices $M$ and that 
the basis on this Hilbert space can be chosen such that a different index
$M(\sigma)$ is attributed to each $\sigma$.
Therefore, the last equation in (\ref{constrLFop}) can be written
in the Schr\"odinger picture as
\begin{equation}\label{KGS1}
(\hat{H}-a)\Psi[X,M]=0 ,
\end{equation}
where $\Psi[X,M]$ is a functional of $X(\sigma)$ and 
$M(\sigma)$, while
\begin{equation}\label{KGS2}
\hat{H}  =  \int d\sigma \left[
\frac{\eta^{\mu\nu}}{2} 
\frac{\delta}{\delta X^{\mu}(\sigma)} \frac{\delta}{\delta X^{\nu}(\sigma)}
- \frac{1}{2} \frac{\partial X^{\mu}(\sigma)}{\partial\sigma} 
\frac{\partial X_{\mu}(\sigma)}{\partial\sigma} 
+\hat{\cal H}_{\rm F} \right] ,
\end{equation}
with
\begin{equation}
\hat{\cal H}_{\rm F}\equiv \frac{i}{2} \eta_{\mu\nu}
\hat{\psi}^{{\sf T}\mu}(\sigma)\rho^0\rho^1\partial_{\sigma}
\hat{\psi}^{\nu}(\sigma) .
\end{equation}
The quantity $a$ can be viewed as a hermitian operator with different eigen-values
on different sectors, making (\ref{KGS1}) correct even for an arbitrary superposition
of states from different sectors. (Recall that the total Hilbert space is the direct sum
of the Hilbert spaces of different sectors.)
The fermionic part $\hat{\cal H}_{\rm F}$ contains the derivative 
$\partial_{\sigma}$, which shows that $\hat{\cal H}_{\rm F}$
vanishes in the pointlike-particle limit. Thus, in this limit,
only the first term in (\ref{KGS2}) survives, leading to the 
Klein-Gordon equation.

Similarly, the supercurrent (\ref{superc}) becomes the operator
\begin{equation}\label{superc2}
\hat{S}_{\alpha}=\frac{1}{2}\rho^0\rho_{\alpha}\hat{\psi}^{\mu}\hat{P}_{\mu}    
+\frac{1}{2}\rho^1\rho_{\alpha}\hat{\psi}^{\mu}
\partial_{\sigma}X_{\mu} .
\end{equation}
Again, the second term vanishes in the pointlike-particle limit 
owing to the derivative $\partial_{\sigma}$. The first term is 
proportional to $\hat{\psi}^{\mu}\hat{P}_{\mu}$, which looks 
similar to the Dirac operator $\gamma^{\mu}\hat{p}_{\mu}$. Indeed, 
comparing (\ref{comrel2}) with (\ref{clif}), we see 
that $\hat{\psi}^{\mu}$ are a sort of stringy analog of the Dirac matrices.
More precisely, in R sector
the relation is of the form \cite{GSW}
\begin{equation}\label{gampsi}
\gamma^{\mu} \propto \int d\sigma \, \hat{\psi}^{\mu}_-(\sigma)=
\int d\sigma \, \hat{\psi}^{\mu}_+(\sigma) .
\end{equation}
Thus, it turns out that the pointlike-particle limit of the 
superstring constraint 
\begin{equation}
\hat{F}_0\Psi[X,M]=0
\end{equation}
is the Dirac equation. However, from (\ref{gampsi}) we see
that, at the fundamental superstring level, $\gamma^{\mu}$ emerges from a 
Lorentz vector.
Similarly, at the fundamental superstring level,
the Dirac wave function $\varphi$ emerges from
a Lorentz scalar $\Psi$. This confirms our 
assertion in Sec.~\ref{SECpfpf} that, at the fundamental level,
$\gamma^{\mu}$ corresponds to a vector and $\varphi$ to a collection of 
scalars.

To avoid confusion, we stress that $\Psi$ and $\varphi$ are scalars
under {\em spacetime} coordinate transformations, but not under
{\em internal} transformations in the Hilbert space with indices $M$.
In particular, the spinor representation of the algebra of
$\gamma^{\mu}$'s corresponds to a tiny subspace of the whole
Hilbert space in which $\hat{\psi}^{\mu}$ live. Thus, this
{\em internal} spinor representation corresponds to the common 
description of fermions in curved spacetime, but
does not play any fundamental role at the unifying superstring level.

Now, by analogy with the particle current studied
in Sec.~\ref{SEC2}, the superstring current
associated with the stringy analog of the Klein-Gordon equation,
Eq.~(\ref{KGS1}), is
\begin{equation}\label{curbosGrS}
J_{\mu}[X;\sigma)=i\int [dM] \,  \Psi^*[X,M]
\frac{
\!\stackrel{\leftrightarrow}{\delta}\! }
{\delta X^{\mu}(\sigma)}
\Psi[X,M] .
\end{equation}
Here the notation $J_{\mu}[X;\sigma)$ denotes  
a functional with respect to $X(\sigma)$ and a function with respect
to $\sigma$, while $[dM]$ denotes the functional
integration over all $M(\sigma)$. 
We see that the superstring current (\ref{curbosGrS}) 
generalizes both (\ref{KGcur}) and (\ref{curbosGr}).
However, the fundamental superstring current (\ref{curbosGrS})
does not require a preferred foliation of spacetime, 
confirming the conjecture in Sec.~\ref{SECpfpf} that the preferred foliation
appears only at the level of effective field theory.
In addition, the superstring current does not depend on the value
of the parameter $a$ (this is analogous to the fact that the
particle current does not depend on the mass $m$), which shows
that the same current can be used for both R and NS sectors.
Finally, since $\hat{\cal H}_{\rm F}$ in (\ref{KGS2}) is a hermitian
operator that acts trivially in the space of functionals of $X$,
and since the dependence on $M$ is integrated out in (\ref{curbosGrS}),
the presence of the term $\hat{\cal H}_{\rm F}$ in (\ref{KGS2})
does not spoil the fact that the superstring current (\ref{curbosGrS}) 
is conserved, in the sense that
\begin{equation}\label{consJ}
\int d\sigma \, \frac{\delta}{\delta X^{\mu}(\sigma)}
J^{\mu}[X;\sigma) = 0.
\end{equation}

Note that $X$ and $M$ in (\ref{curbosGrS}) and (\ref{consJ}) are not treated on an
equal footing, which means that the world-sheet supersymmetry is not manifest.
However, the world-sheet supersymmetry should be distinguished from the
spacetime supersymmetry. The spacetime bosons and fermions
are unified in the sense that the same current $J^{\mu}$ describes both
spacetime bosons and spacetime fermions.  

\section{The Bohmian interpretation}
\label{BOHM}

\subsection{Bohmian interpretation of particles}
\label{BOHM1}

What is the physical interpretation of the current (\ref{KGcur})
associated to the wave function (\ref{wffield})?
As the time component $j_0(x)$ may be negative, it cannot be interpreted
as the probability density of particle positions. The time component
cannot be interpreted as the charge density either, because
the hermitian field $\hat{\phi}$ cannot describe a charged particle.
A viable interpretation (see, e.g., \cite{durpra1,nikfpl3}) is 
the Bohmian interpretation, according to which 
the current determines the trajectory of the particle $X^{\mu}(\tau)$
according to the law
\begin{equation}\label{bohm1}
\frac{dX^{\mu}}{d\tau}=j^{\mu}(X) ,
\end{equation}
where $\tau$ is an affine parameter along the trajectory. 
Thus, the trajectory is nothing but an integral curve of the vector field
$j^{\mu}(x)$. As such, the trajectory does not depend on the 
parametrization with $\tau$. The reparametrization of all trajectories
in the congruence of integral curves is equivalent to a local rescaling 
of the current
\begin{equation}\label{rescal}
j^{\mu}(x) \rightarrow j'^{\mu}(x)=e^{\Omega(x)}j^{\mu}(x) 
\end{equation}
in Eq.~(\ref{bohm1}), where $\Omega(x)$ is an arbitrary
real function. If $j^{\mu}$ is conserved, i.e., if 
$\partial_{\mu}j^{\mu}=0$, then $j'^{\mu}$ may not be conserved:
\begin{equation}\label{nonconsb}
\partial_{\mu}j'^{\mu}=j'^{\mu}\partial_{\mu}\Omega .
\end{equation}
Thus, for the consistency of the Bohmian interpretation
(\ref{bohm1}), the current $j^{\mu}$ does not necessarily need 
to be conserved.

By writing $\varphi=Re^{iS}$, where $R$ and $S$ are real functions,
the current (\ref{KGcur}) is equal to $j_{\mu}=-2R^2\partial_{\mu} S$.
By a local rescaling this can be transformed to
$j_{\mu}=-\partial_{\mu} S$. Therefore, the equation of motion
(\ref{bohm1}) can be written in the Hamilton-Jacobi form as
\begin{equation}\label{bohm2}
\frac{dX^{\mu}}{d\tau}=-\partial^{\mu} S(X) .
\end{equation}
However, the existence of such a Hamilton-Jacobi form is a 
consequence of the special form of the spin-0 current 
(\ref{KGcur}). For other spins, such a Hamilton-Jacobi form
for the particle equation of motion does not exist.
In particular, for the current (\ref{curbos}) where
$\varphi$ is a many-component wave function, a Hamilton-Jacobi form
does not exist. Instead, for arbitrary spin, the motion of the 
particle is described by an equation of the form of 
(\ref{bohm1}) with an appropriate particle current $j^{\mu}$.

Now let us generalize this to the many-particle case.
For simplicity, consider a many-particle wave function 
$\varphi(x_1,\ldots,x_n)$ describing $n$ particles
with spin 0. Thus, there are $n$ particle currents 
$j^{\mu}_a=i \varphi^* \!\stackrel{\leftrightarrow\;}{\partial^{\mu}_a}\!
\varphi$, $a=1,\ldots,n$, 
generalizing the current (\ref{KGcur}). Each current is conserved, i.e.,  
$\partial_{a\mu}j^{\mu}_a =0$ for each $a$. Consequently, there is also
an overall conservation equation 
\begin{equation}\label{consoverall}
\sum_a \partial_{a\mu}j^{\mu}_a =0 .
\end{equation}
Therefore, the Bohmian equation
of motion (\ref{bohm1}) generalizes to $n$ coupled equations \cite{nikfpl3}
\begin{equation}\label{bohm1n}
\frac{dX^{\mu}_a}{d\tau}=j_a^{\mu}(X_1,\ldots,X_n) ,
\end{equation} 
for $n$ trajectories $X^{\mu}_a(\tau)$. For each $\tau$, the right-hand 
side of (\ref{bohm1n}) depends on the positions of {\em all}
particles at the same $\tau$. In other words, the velocity of one 
particle depends on the ``instantaneous" position of all other particles.
Such nonlocal ``instantaneous" communication is exactly what is needed
to make the nonlocal correlations typical of QM \cite{bell} consistent
with the notion of objective reality. 
Nevertheless, we see that the equation of motion (\ref {bohm1n}) is
relativistic covariant, i.e., this equation does not depend on the choice
of the preferred foliation. 
By a local rescaling, (\ref{bohm1n}) can also be written in a form generalizing
(\ref{bohm2}) 
\begin{equation}\label{bohm2n}
\frac{dX^{\mu}_a}{d\tau}=-\partial^{\mu}_a S(X_1,\ldots,X_n) .
\end{equation}

A remarkable property of the currents such as 
(\ref{KGcur}) and (\ref{curbos}) is that these vector fields may
be spacelike at some regions of spacetime, even when
$\varphi$ is a superposition of positive-frequency solutions only.
Consequently, the particle velocity (\ref{bohm1}) can exceed 
the velocity of light. 
As shown in \cite{nikfpl1}, this does not contradict experiments
because a {\em measured} velocity, associated to the eigenvalues
of the velocity (or momentum) operator, can never exceed the 
velocity of light.

\begin{figure*}[t]
\includegraphics[width=15.1cm]{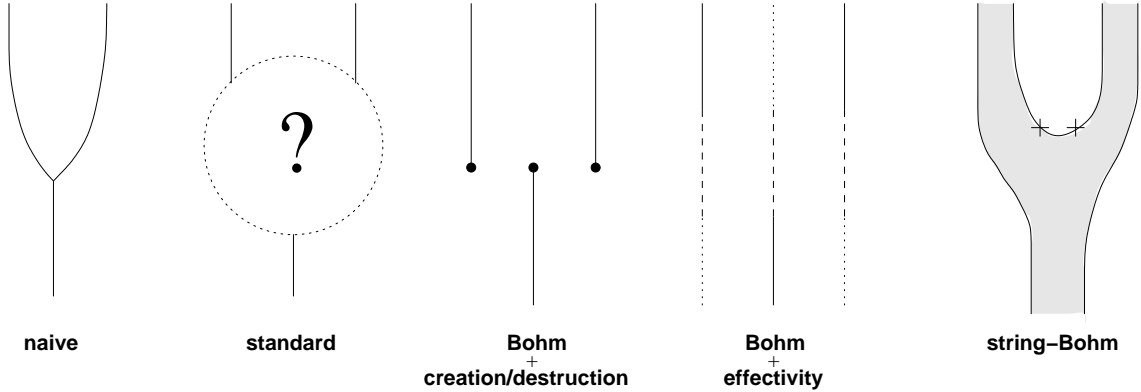}
\caption{\label{fig1}
Various $1+1$ spacetime pictures of a decay of one particle
into two particles.
In a {\em naive} picture (which should not be confused with a similarly
looking Feynman diagram), the particle literally splits into
two particles, leading to a singular splitting point. No reasonable actual
theory supports such a picture.
In the {\em standard} S-matrix picture,
only the measured asymptotic initial and final states are known,
providing no answer to the question what happens at intermediate times
at which measurements are not performed.
In the {\em Bohm+creation/destruction} picture, particles
suddenly and stochastically get created or destructed at singular points.
In the {\em Bohm+effectivity} picture, particle trajectories
never get created or destructed. Solid parts of trajectories correspond to
fully effective particles with $e=1$, dashed parts correspond to
$1>e>0$, while dotted parts correspond to
ineffective (``virtual") particles with $e=0$.
In the {\em string-Bohm} picture (which, again, should not be
confused with a similarly looking Feynman diagram),
the apparent string creation and destruction actually corresponds
to a continuous distortion of a single string. The string boundary
between the two crosses moves faster than light.
Note that the most advanced string-Bohm picture is 
conceptually the most similar
to the ``naive" one, but contains no singular splitting point.}
\end{figure*}

There are also qualitative suggestions that superluminal 
velocities with associated motions backwards in time 
may be related to the physical creation and destruction of particles.
However, such superluminal motions appear even for free particles 
corresponding to the free Klein-Gordon equation (\ref{KG}),
whereas the physical creation and destruction requires 
field interactions. Therefore, to make the existence of 
Bohmian particle trajectories consistent with the standard
QFT predictions on particle creation and destruction, it turns out that
one is forced to introduce some additional structure
that, unfortunately, makes the theory less elegant.
One possibility is to postulate stochastically chosen 
singular points at which
particle trajectories begin or end, corresponding to particle
creation or destruction \cite{durrjpa,durrprl}.
To avoid singular points and stochastic processes that break the spirit
of the Bohmian interpretation, another possibility is to 
introduce an additional continuously evolving property of the
particle called {\em effectivity} \cite{nikfpl1,nikfpl2}, which leads to
a picture in which particle trajectories never begin or end.
Instead, the creation corresponds to a process in which 
the effectivity $e$ changes continuously from  $e=0$ to $e=1$,
while the destruction corresponds to a similar continuous fading
from $e=1$ to $e=0$. Admittedly, both possibilities seem
somewhat artificial. However, as we shall see in the next subsection,
the Bohmian interpretation of strings offers a much more 
elegant picture of the processes of particle creation and destruction.
Moreover, superluminal velocities and associated motions backwards
in time that may seem undesirable at the particle level turn out 
to play an appealing role in the Bohmian description
of string creation and destruction, thus reinforcing the viability
of particle currents that may lead to superluminal motions.
Various pictures of particle
creation and destruction, including the 
Bohmian string picture studied in the next section, 
are illustrated in Fig.~\ref{fig1}.     

\subsection{Bohmian interpretation of strings}

Let us start with the Bohmian interpretation of bosonic 
(rather than supersymmetric) string theory.
In the bosonic case, the term $\hat{{\cal H}}_{\rm F}$ in 
(\ref{KGS2}) is absent. Consequently, there are no $M$-indices
and the string wave functional is of the form $\Psi[X]=R[X]e^{iS[X]}$, 
where $R$ and $S$ are real. The Bohmian interpretation introduces
a deterministic evolution of the string coordinates
$X^{\mu}(\sigma,\tau)$, given by
an equation of motion having a Hamilton-Jacobi form \cite{nikstring1,nikdual}
\begin{equation}\label{bohm2s}
\frac{\partial X^{\mu}(\sigma,\tau)}{\partial\tau}=-\eta^{\mu\nu} 
\frac{\delta S[X]}{\delta X^{\nu}(\sigma)} ,
\end{equation}
which is analogous to (\ref{bohm2}). In (\ref{bohm2s}), $\sigma$ and $\tau$
play different roles, which breaks the manifest world-sheet covariance
of the Bohmian interpretation. However, there is a way to write
(\ref{bohm2s}) in a world-sheet covariant form \cite{nikstring1},
with a dynamically generated preferred foliation of the world-sheet
({\em not} of spacetime!).
Another way to write (\ref{bohm2s}) is
\begin{equation}\label{stringbohm1}
\frac{\partial X^{\mu}(\sigma,\tau)}{\partial\tau}= \frac{1}{2R^2[X]} 
J^{\mu}[X;\sigma) ,
\end{equation}
where
\begin{equation}\label{curbosGrSb}
J_{\mu}[X;\sigma)=i\Psi^*[X]
\frac{
\!\stackrel{\leftrightarrow}{\delta}\! }
{\delta X^{\mu}(\sigma)}
\Psi[X] .
\end{equation} 
Analogously to that in Sec.~\ref{BOHM1}, the parameter $\tau$ can be
redefined so that the factor $1/2R^2$ in (\ref{stringbohm1}) gets 
eliminated, leading to the equation analogous to (\ref{bohm1})
\begin{equation}\label{stringbohm2}
\frac{\partial X^{\mu}(\sigma,\tau)}{\partial\tau}= J^{\mu}[X;\sigma) .
\end{equation}

At this point it is fair to remark that the string Bohmian equations of motion
different from (\ref{bohm2s}) are also conceivable.
Nevertheless, there are several theoretical arguments for preferring
(\ref{bohm2s}). First, (\ref{bohm2s}) is the obvious generalization
of the particle Bohmian equation of motion (\ref{bohm2}).
Second, (\ref{bohm2s}) represents the obvious generalization
of classical string dynamics, as the classical Hamilton-Jacobi
string equation of motion takes the same form as (\ref{bohm2s}) \cite{nikstring1}.
Third, the equation of motion (\ref{bohm2s}) can be derived from 
the world-sheet covariant canonical quantization of strings introduced in \cite{nikstring1}, where the Bohmian equation of motion is {\em derived} (not simply postulated for interpretational purposes)
from the requirement that the world-sheet covariant canonical quantization based on the De Donder-Weyl formalism should be compatible with the standard
canonical quantization.
Fourth, (\ref{bohm2s}) has a property of manifest spacetime relativistic covariance.
Some additional arguments for preferring (\ref{bohm2s}) will be presented in Sec.~\ref{PROB}.
%


In \cite{nikstring1,nikdual}, the Bohmian interpretation was studied
for free bosonic strings only. However, one of the
most remarkable properties of perturbative string theory
(i.e., theory defined by the perturbative expansion over Feynman diagrams
with different world-sheet topologies)
is the fact that the string analog of particle creation and destruction
is also described by the {\em free} string Lagrangian. (The specific 
interactions of the low-energy field-theoretic 4-dimensional Standard Model
are encoded in a peculiar relation between 
the observed 4 dimensions and the unobserved extra dimensions
of string theory.) 
Thus, the free string wave functional $\Psi[X]$ also contains 
the information about the probabilities for string creation and destruction.
To see how, the crucial observation is that
the functional $\Psi[X]$ may be nonvanishing even for 
functions $X^{\mu}(\sigma)$ that are not continuous. 
For example, a discontinuous function of the form
\begin{equation}
X^{\mu}(\sigma)=\left\{ 
\begin{array}{c}
X_1^{\mu}(\sigma) \;\;\;{\rm for} \;\;\; 
  0\leq\sigma<\displaystyle\frac{\pi}{2} , \\
X_2^{\mu}(\sigma) \;\;\;{\rm for} \;\;\; 
  \displaystyle\frac{\pi}{2}<\sigma\leq{\pi} ,    
\end{array}
\right.
\end{equation}
where $X_1^{\mu}(\sigma)$ and $X_2^{\mu}(\sigma)$ are continuous
functions, may be viewed as a configuration that describes
{\em two} strings. Consequently, for such functions,
one can write $\Psi[X]=\Psi[X_1,X_2]$, describing a state with two strings.
For entangled strings corresponding to the case
$\Psi[X]=\Psi[X_1,X_2]\neq \Psi_1[X_1]\Psi_2[X_2]$, the Bohmian
interpretation requires a preferred foliation of the world-sheet
(which can be introduced dynamically \cite{nikstring1}), but 
a preferred foliation of spacetime is not needed here.
Nevertheless, a relation with the foliation of spacetime exists:
the vector $n^{\mu}(x)$ associated with the foliation of spacetime  
can be projected onto the world-sheet, which then defines 
a foliation of the world-sheet as well. This does not 
uncover a more fundamental origin of the preferred foliation, 
but suggests that Bohmian nonlocality and string theory
may be related in a yet undiscovered way. (For an overview
of nonlocal phenomena emerging from string theory, see \cite{nl1,nl2},
and for a more stringent relation with Bohmian nonlocalities,
see \cite{nikdual}.)

To see more explicitly how the description of string creation and
destruction in the Schr\"odinger picture is related to the usual
formulation of string theory \cite{GSW,polc}, we proceed as follows.
We start from the functional Schr\"odinger equation
\begin{equation}\label{schstr}
\hat{H}\Psi[X;\tau)=i\partial_{\tau}\Psi[X;\tau) ,
\end{equation}
where, for convenience, the constant $a$ is absorbed into a 
redefinition of the Hamiltonian $\hat{H}$. As is well known
from particle quantum mechanics and QFT \cite{ryder}, 
the evolution described by the Schr\"odinger equation is equivalent to 
\begin{equation}\label{propstr1}
\Psi[X;\tau)=\int [dX'] \, \langle X;\tau|X';\tau_0\rangle \Psi[X';\tau_0) ,
\end{equation}
where $[dX]$ is a functional integral with respect to all
$X^{\mu}(\sigma)$, the propagator
$\langle X;\tau|X';\tau_0\rangle$ is given by the path integral
\begin{equation}\label{path1}
\langle X_A;\tau_A|X_B;\tau_B\rangle =
\int_{X_A(\sigma),\tau_A}^{X_B(\sigma),\tau_B} [dX(\sigma,\tau)] \,
e^{iA[X]} ,
\end{equation}
and $A$ is the classical action (\ref{Act}) (without the fermionic 
part depending on $\psi^{\mu}$, as we deal here with  
bosonic string theory).
In the path integral above, various functions $X^{\mu}(\sigma,\tau)$
may be discontinuous at various points and 
can be classified according to different topologies of the corresponding 
world-sheets. Therefore, (\ref{path1}) can be further written as
\begin{equation}\label{path2}
\langle X_A;\tau_A|X_B;\tau_B\rangle = \sum_g
\int_{X_A(\sigma),\tau_A}^{X_B(\sigma),\tau_B} [dX(\sigma,\tau)]_g \,
e^{iA[X]} ,
\end{equation}
where $[dX(\sigma,\tau)]_g$ denotes the integration over functions that
describe the same topology.
So far we have been assuming that $\Psi$ satisfies the
Schr\"odinger equation (\ref{schstr}), but actually
this is not the only equation
that $\Psi$ satisfies. First, it also satisfies the Hamiltonian constraint
$\hat{H}\Psi=0$, which means that $\Psi$ in (\ref{schstr}) does not
really depend on $\tau$, and consequently, that the propagators
(\ref{path1}) and (\ref{path2}) do not really depend on
$\tau_A$ and $\tau_B$. Second, it also satisfies all other
Virasoro constraints for $n>0$ (see Eq.~(\ref{constrLFop})), 
which implies that the propagator does not depend on the choice 
of the world-sheet metric $h_{\alpha\beta}(\sigma,\tau)$. 
Therefore, in (\ref{path2}),
one must replace the implicit Minkowski world-sheet metric
$\eta_{\alpha\beta}$ in (\ref{Act}) by an arbitrary metric
$h_{\alpha\beta}(\sigma,\tau)$ and integrate over all possible 
metrics. This leads to
\begin{equation}\label{path3}
\langle X_A|X_B\rangle = \sum_g
\int_{X_A(\sigma)}^{X_B(\sigma)} [dh(\sigma,\tau)]_g [dX(\sigma,\tau)]_g \, 
e^{iA[X,h]} ,
\end{equation}  
where $[dh(\sigma,\tau)]_g$ denotes the functional integral over all possible
world-sheet metrics $h_{\alpha\beta}(\sigma,\tau)$
consistent with given topology of the world-sheet. 
(In the bosonic case, (\ref{path3}) turns out to be well defined only for
$D=26$.) Consequently, (\ref{propstr1}) is replaced by
\begin{equation}\label{propstr2}
\Psi[X]=\int [dX'] \, \langle X|X'\rangle \Psi[X'] .
\end{equation}
Eq.~(\ref{path3}) is nothing but the usual path-integral 
formula for calculating the string scattering amplitudes, 
where the summation over $g$ corresponds to the summation over
Feynman diagrams with different topologies.
The quantity $\Psi[X]$ in (\ref{propstr2}) is the string wave functional
that defines the bosonic string current (\ref{curbosGrSb}). 

As is well known from the Bohmian interpretation of particles and 
fields, solutions of the Bohmian equation of motion (\ref{bohm2s})
can only attain configurations for which the 
amplitude 
of 
$\Psi[X]$ does not vanish. 
Moreover, when the quantum-mechanical probabilities
are well-defined by $\Psi[X]$, then the corresponding probabilities
predicted by the Bohmian interpretation turn out to be exactly the same.
(We discuss it in more detail in Sec.~\ref{PROB}.)
Now, since $\Psi[X]$ contains amplitudes corresponding to 
different world-sheet topologies, including those that correspond 
to string splitting, it is evident that the Bohmian equation of motion
(\ref{stringbohm2}) contains solutions corresponding to the 
same world-sheet topologies. In particular, some solutions  
describe deterministic processes of string splitting, which
correspond to particle creation and destruction (see Fig.~\ref{fig1}
for an example). In fact, such splitting solutions exist even in
classical string theory \cite{sanchez}.
In this way, unlike the Bohmian interpretation
of pointlike particles, the Bohmian interpretation of strings 
does not need an artificial introduction of stochastic 
singular points at which the particle trajectories begin and end, 
or an artificial introduction of effectivities. Thus,
the Bohmian interpretation of strings provides a very elegant solution
to the problem of the Bohmian description of particle creation and destruction
(see also \cite{nikisqftq}). 

Now we also see why it is appealing that the current such as
(\ref{curbosGrSb}) may be locally spacelike. In the Bohmian interpretation
this corresponds to local superluminal velocities, which, as 
demonstrated by Fig.~\ref{fig1}, is related to string creation and
destruction without leading to singular splitting points, 
provided that $S[X]$ is sufficiently smooth, leading to a 
smooth right-hand side of (\ref{bohm2s}).
(Without superluminal velocities, the {\sf U}-shape of the string boundary 
between the two crosses would be replaced by a {\sf V}-shape 
containing a singular splitting point at the cusp.) 

Now the generalization of the bosonic-string results above 
to superstrings is 
straightforward, as we now discuss briefly.
It is evident that the superstring generalization of (\ref{curbosGrSb})
is the superstring current (\ref{curbosGrS}). Thus, the natural
interpretation of this superstring current is the Bohmian interpretation
defined by (\ref{stringbohm2}), i.e., 
\begin{equation}\label{stringbohm2.2}
\frac{\partial X^{\mu}(\sigma,\tau)}{\partial\tau}= J^{\mu}[X;\sigma) .
\end{equation}
Note that, in the superstring case, 
there is no Bohmian equation of motion for the ``classical"
Grassmann valued coordinate $\psi^{\mu}$ in (\ref{Act}). 
This is analogous to the fact that in particle Bohmian mechanics
there is no Bohmian equation of motion for the spin of the particle
\cite{durrnaive}.
The spin degrees of freedom are integrated out in the particle
case (\ref{curbosGr}) in the same manner as the degrees of freedom
attributed to $\hat{\psi}^{\mu}$ are integrated out in the 
superstring case (\ref{curbosGrS}).
In this way, the Bohmian interpretation of superstrings provides
a sort of explicit realization of the idea that in Bohmian mechanics 
``all particles are identical" \cite{durrall},
i.e., that all particles are merely different states of the same particle
(which here turns out to be an extended ``particle'', that is -- string). 

\section{Probabilistic predictions and consistency with the Bohmian interpretation}
\label{PROB}

In the previous section we have presented the deterministic Bohmian equations
of motion for particles and strings. However, we have not explicitly explained
how the usual probabilistic predictions of the conventional formulation
of quantum theory can be recovered from these deterministic 
equations of motion. This is what we do in this section.

Our approach is based on the old idea \cite{stuec,horw,kypr,fanchi}
that in relativistic QM the quantity $|\varphi(x)|^2$ represents the probability
density of the particle position in {\em spacetime}, rather than that in space.
Such an approach has an advantage that the probabilistic interpretation
is manifestly relativistic
covariant and can be easily generalized to  
many particles and strings, without introducing any preferred foliation of
spacetime. Here, partially inspired by the results of \cite{durpra1}, 
we construct a Bohmian version of that interpretation.
Actually, we further refine the ideas of  \cite{durpra1} by showing 
that, even with the Bohmian interpretation,
the probabilistic predictions of particle positions 
are easily found on {\em any} hypersurface.
(For further developments see also \cite{niktime,nikQFT}.)

The simplest way to understand our approach is to exploit the analogy
with the well-understood case of free nonrelativistic particles in an
energy eigenstate, by discussing the latter from a somewhat unusual point of view.
Thus, the first part of this subsection is devoted to a detailed discussion
of the nonrelativistic case, while the most interesting relativistic case is then
presented in the second part as a simple generalization.

\subsection{Nonrelativistic particles in an energy eigenstate}

A free nonrelativistic particle in an energy eigenstate is described by a 
wave function of the form $\psi(t,{\bf x})=e^{-iEt}\varphi({\bf x})$, where 
$ {\bf x}\equiv \{x^1,x^2,x^3\}$.
Thus, the nonrelativistic
Schr\"odinger equation reduces to the time-independent Schr\"odinger equation
$-\nabla^2\varphi/2m =E\varphi$, which we write as
\begin{equation}\label{pr1}
 (\nabla^2+M^2)\varphi({\bf x})=0,
\end{equation}
where $M^2\equiv 2mE$. We write the time-independent Schr\"odinger equation
in the form (\ref{pr1}) to make the analogy with the relativistic Klein-Gordon equation
(\ref{KG}) obvious. The current 
\begin{equation}\label{pr2}
{\bf j}=\frac{ i \varphi^* \!\stackrel{\leftrightarrow\;}{\nabla}\! \varphi }{2m}
\end{equation}
satisfies the space-conservation equation
\begin{equation}\label{pr3}
 \nabla {\bf j}=0 ,
\end{equation}
which are (up to an inessential normalization factor $1/2m$)
nonrelativistic analogs of (\ref{KGcur}) and (\ref{KGcur'}), respectively.
How to interpret (\ref{pr3}) in terms of conserved probabilities?
Since $\varphi$ does not depend on $t$, (\ref{pr3}) also trivially implies 
a conservation equation involving time
\begin{equation}\label{pr4}
\frac{ \partial |\varphi|^2}{\partial t} + \nabla (|\varphi|^2 {\bf v})=0 ,
\end{equation}
where ${\bf v}\equiv{\bf j}/|\varphi|^2=\nabla S/m$ and $\varphi=Re^{iS}$.
Thus, it is consistent to postulate that the {\it a priori} probability density
of particle positions is
\begin{equation}\label{pr5}
 p({\bf x})=|\varphi({\bf x})|^2 .
\end{equation}
Consequently, it is also consistent to postulate that particles have trajectories
{\bf X}(t) satisfying
\begin{equation}\label{pr6}
 \frac{d{\bf X}}{dt}=\frac{{\bf \nabla}S}{m} ,
\end{equation}
which is a nonrelativistic analog of (\ref{bohm2}) (up to an additional inessential rescaling of $\tau$ in (\ref{bohm2})). Indeed, (\ref{pr6}) is the standard nonrelativistic 
Bohmian equation of motion. 

In practice, one rarely measures the probability density  (\ref{pr5}) in the whole space.
Instead, one frequently deals with a beam moving in the $x^3$ direction and measures
particle positions on the detection screen at $x^3=z$. It is convenient 
to model the beam by a wave function localized in the $x^1$-$x^2$ directions, 
but infinitely extended in the $x^3$ direction. For example, the wave function 
may have the form
\begin{equation}\label{pr6.1}
\varphi({\bf x})=e^{ik^3x^3}\chi(x^1,x^2), 
\end{equation}
where $\chi(x^1,x^2)$
is a localized wave packet, but a more general dependence on $x^3$ is also possible. 
Thus we have
\begin{equation}\label{pr7}
 \int_{-\infty}^{\infty} dx^3 |\varphi(x^1,x^2,x^3)|^2 = \infty .
\end{equation}
Nevertheless, the property (\ref{pr7}) does not represent a practical problem, 
because the measured probability density of particle positions at $x^3=z$ is
\begin{equation}\label{pr8}
p_z(x^1,x^2)=\frac{|\varphi(x^1,x^2,z)|^2}{N_z} ,
\end{equation}
where
\begin{equation}\label{pr9}
N_z=\int dx^1 dx^2 |\varphi(x^1,x^2,z)|^2
\end{equation}
is the normalization factor. Even though the wave function cannot be normalized
in the whole space due to (\ref{pr7}), the physically relevant normalization factor
(\ref{pr9}) is finite. The probability density (\ref{pr8}) is nothing but the conditional
probability density derived from the joint probability density (\ref{pr5}). 

It is well known that (\ref{pr8}) is consistent with the Bohmian interpretation,
but it is crucial to understand why exactly this is so. The Bohmian trajectories
are given by 3 functions 
\begin{equation}\label{pr10}
X^1(t), \; X^2(t), \; X^3(t) 
\end{equation}
satisfying (\ref{pr6}).
Now assume that an experimentalist is {\em not} equipped with a clock
that measures $t$. In particular, he does not measure the time at which the 
particle approaches
the detection screen at $x^3=z$. Thus, for practical purposes, 
{\em the parameter 
$t$ can be thought of as an auxiliary mathematical parameter without a direct physical
interpretation.} Hence, it is convenient to eliminate $t$ from (\ref{pr10}).
From the last function in (\ref{pr10}) one can find $t(X^3)\equiv t(x^3)$ and 
insert this into the first two
functions. In this way one ends up with only two functions 
\begin{equation}\label{pr11}
X^1(x^3), \; X^2(x^3). 
\end{equation}
Geometrically, (\ref{pr10}) is a curve in 3+1=4 dimensions, while (\ref{pr11})
is a curve in only 2+1=3 dimensions. In fact, and this is the crucial point,
{\em the physical coordinate $x^3$ in (\ref{pr11}) plays a role of a time coordinate}.
In this picture, $\psi(x^1,x^2,x^3)$
describes how the wave function evolves with 
``time'' $x^3$, while (\ref{pr8}) is the probability density on a surface
of constant ``time'' $x^3=z$.  The crucial point to note is that, with the dependence
on $x^3$ more general than that in (\ref{pr6.1}), the norm (\ref{pr9})
depends on $z$. Consequently, the probability density on a surface 
of constant ``time'' is not simply given by the numerator 
on the right-hand side of (\ref{pr8}).
Nevertheless, the probability density {\em is} proportional to this numerator,
with the proportionality factor being $z$-dependent as expressed by 
the denominator on the right-hand side of (\ref{pr8}).
 
How can (\ref{pr8}) be understood from the Bohmian 2+1 point of view?
In general, the trajectories (\ref{pr11})
may have a very strange appearance from the 2+1 perspective. 
For example, by allowing a more complicated dependence on $x^3$ in 
(\ref{pr6.1}), the trajectories may even
involve motions backwards in ``time''. Nevertheless, (\ref{pr8}) is the correct
probability density on $x^3=z$. This is because the {\it a priori} probability
density on the whole space is given by (\ref{pr5}). From the 2+1 perspective,
it means that not only the initial particle position $X^1(x^3_{\rm initial})$,
$X^2(x^3_{\rm initial})$ is unknown and hence subject to the probabilistic law, 
but also the value of the initial
``time'' $x^3_{\rm initial}$ itself is unknown and hence subject to the probabilistic law.

Now let us briefly generalize all this to an $n$-particle energy eigenstate.
Most of it is obvious, so let us only discuss some less obvious features.
The $n$-particle energy eigenstate
is described by a wave function $\varphi({\bf x}_1, \ldots, {\bf x}_n)$.
The {\it a priori} probability density (\ref{pr5}) generalizes to
\begin{equation}\label{pr5.1}
 p({\bf x}_1, \ldots, {\bf x}_n)=|\varphi({\bf x}_1, \ldots, {\bf x}_n))|^2 .
\end{equation}
Since it does not depend on $t$, it is the correct joint probability density
even if different particles are detected at different $t$.
(Note that the calculation of the joint probability density does not involve 
an effective collapse of the wave function. The collapse is only related
to the calculation of a conditional probability density, such as calculating 
the probability that the particle 2 has the position ${\bf x}_2$ at time
$t_2$, given that the particle 1 has been found at the position ${\bf x}_1$ at time
$t_1$.)  
As in the 1-particle case above, we are interested in a situation in which
$t$ is an unmeasured (and hence effectively unphysical) parameter.
In general,
it can be arranged that different particles are detected on screens 
positioned at different values of $x^3$. Hence, (\ref{pr8})-(\ref{pr9}) generalize to 
\begin{equation}\label{pr8.1}
p_{z_1,\ldots,z_n}({\bf x}^{\perp}_1,\ldots,{\bf x}^{\perp}_n)
=\frac{|\varphi({\bf x}^{\perp}_1,z_1,\ldots, {\bf x}^{\perp}_n,z_n)|^2}
{N_{z_1,\ldots,z_n}} ,
\end{equation}
\begin{equation}\label{pr9.1}
N_{z_1,\ldots,z_n}=\int d^2{\bf x}^{\perp}_1 \ldots \int d^2{\bf x}^{\perp}_n
|\varphi({\bf x}^{\perp}_1,z_1,\ldots, {\bf x}^{\perp}_n,z_n)|^2 ,
\end{equation}
where ${\bf x}^{\perp}_a \equiv \{ x^1_a,x^2_a \}$.
The Bohmian trajectories ${\bf X}_a(t)$ can be viewed as $n$ curves in 4 dimensions.
But time can be eliminated, leading to functions ${\bf X}^{\perp}_a(x^3_a)$, 
which can be viewed as $n$ curves in 3 dimensions ${\bf X}^{\perp}_a(x^3)$. 
Now (\ref{pr8.1}) is consistent with the trajectories ${\bf X}^{\perp}_a(x^3)$
because not only the initial conditions 
${\bf X}^{\perp}_a(x_{a\;{\rm initial}}^3)$
are subject to the probabilistic law (\ref{pr5.1}), 
but so are the initial ``times'' $x_{a\;{\rm initial}}^3$
themselves. 

\subsection{Relativistic particles and strings}

Now we are finally ready to deal with the relativistic case. Actually, with the
nonrelativistic case being described as above, the generalization to the
relativistic case is very simple. Essentially,
all we need to do is to add one dimension more.
More precisely, the transition from a nonrelativistic particle in an energy eigenstate
to a relativistic particle is accomplished through the following correspondence:
\begin{eqnarray}\label{coresp}
& t\rightarrow \tau , & \nonumber \\
& {\bf x}=(x^1,x^2,x^3) \rightarrow x=(x^0,x^1,x^2,x^3)  ,  & \nonumber \\
& {\bf x}^{\perp}=(x^1,x^2) \rightarrow {\bf x}=(x^1,x^2,x^3)  , 
\;\;\;\; x^3 \rightarrow x^0  . & 
\end{eqnarray}
In particular, the parameter $\tau$ in (\ref{bohm2}) can be thought of as a 
scalar time corresponding to a fifth dimension \cite{stuec,horw,kypr,fanchi}.
However, $\tau$ is an auxiliary unphysical parameter, i.e., we are not equipped with a clock that
measures it. Hence it is natural to eliminate $\tau$, so that (\ref{bohm2}) describes
a curve in 4 physical dimensions. The wave function $\varphi(x)$ does not depend
on $\tau$, so (\ref{pr4}) generalizes to
\begin{equation}\label{pr4.2}
 \frac{\partial |\varphi|^2}{\partial\tau} + \partial_{\mu} (|\varphi|^2 v^{\mu})=0 ,
\end{equation}
where $v^{\mu}\equiv j^{\mu}/|\varphi|^2=-\partial^{\mu} S$
and the second term in (\ref{pr4.2}) vanishes due to (\ref{KGcur'}).
Consequently, (\ref{pr5}) generalizes to
\begin{equation}\label{pr5.2}
 p(x)=|\varphi(x)|^2 ,
\end{equation}
which defines a probability density conserved in $\tau$. Since $\tau$ is unphysical,
(\ref{pr5.2}) can also be thought of as the {\it a priori} probability density
that particle will be found at any point $x$ in spacetime. In other words, 
(\ref{pr5.2}) is the probability density of {\em flashes} in spacetime.

At this point it is interesting to note that flashes in spacetime have also been
introduced in a different context \cite{tumulka}, as a way to make the 
objective-collapse interpretation of QM fully compatible with relativity.
In our approach, (\ref{pr5.2}) will be the ultimate reason why the probabilistic
predictions of the Bohmian interpretation are fully compatible with relativity.
Thus, in both cases the ultimate origin of the compatibility between relativity
and hidden variables for QM lies in a use of flashes in spacetime,
which cannot be a coincidence. Nevertheless, in our Bohmian approach 
the particles are not flashes but continuous trajectories in spacetime, while the flashes 
correspond only to initial conditions. Namely, a Bohmian trajectory can be specified
by fixing $X^{\mu}(\tau=0)$, which, after elimination of the variable $\tau$,
means that one must fix $X^i(x^0_{\rm initial})$, $i=1,2,3$, as well as
$x^0_{\rm initial}$ itself. The {\it a priori} probability density for such an
initial condition is given by (\ref{pr5.2}).

The probability density (\ref{pr5.2}) cannot be normalized in the whole spacetime,
due to an analog of (\ref{pr7}) 
\begin{equation}\label{pr7.2}
 \int dx^0 |\varphi(x^0,x^1,x^2,x^3)|^2 = \infty .
\end{equation}
Nevertheless, this is not a problem in practice if one is interested in the probability
for a fixed $x^0$. If one is certain that a particle will be detected at 
the time $x^0=t$, then the probability distribution of particle positions
in space at $x^0=t$ is given by the conditional probability density
\begin{equation}\label{pr8.2}
p_t({\bf x})=\frac{|\varphi(t,{\bf x})|^2}{N_t} ,
\end{equation}
where 
\begin{equation}\label{pr9.2}
N_t=\int d^3x |\varphi(t,{\bf x})|^2 .
\end{equation}
These equations are nothing but obvious generalizations of (\ref{pr8}) and  (\ref{pr9}).
In nonrelativistic QM, where $\varphi(t,{\bf x})$ satisfies a first-order differential
equation with respect to $t$, the norm (\ref{pr9.2}) does not depend on $t$.
In relativistic QM described by the Klein-Gordon equation this norm may depend
on $t$ (actually, it depends on $t$ only if $\varphi$ is a superposition 
of solutions with both positive and negative frequencies \cite{nikprobbohm}), 
which is the standard argument  that $|\varphi|^2$ is not the probability density
at fixed time
in relativistic QM. Nevertheless, now we see that the probability density at fixed time
is proportional to $|\varphi|^2$, with the time-dependent proportionality factor
given by the denominator on the right-hand side of (\ref{pr8.2}). 

Given the {\it a priori} probability density of flashes (\ref{pr5.2}),
it is clear that (\ref{pr8.2}) is the correct conditional probability density.
However, in nonrelativistic QM, (\ref{pr8.2}) (with constant $N_t$)
is interpreted as an {\it a priori} probability density, i.e., one allways assumes
that one {\em is} certain that the particle will be detected at any fixed value of $t$.
Can one be certain that this will also be the case when one starts from
(\ref{pr5.2})? While it may be a difficult conceptual question within the 
orthodox probabilistic interpretation of QM (which may be a reason why the
probabilistic interpretation of QM advocated in  \cite{stuec,horw,kypr,fanchi}
is not widely accepted), this question has a simple answer within the Bohmian
interpretation. Since particles are actually trajectories (not merely flashes)
in spacetime, one expects that every trajectory associated with a free 
Klein-Gordon equation crosses any given hypersurface of
equal time. Indeed, we have performed some numerical computations
that give evidence that 
this is indeed the case when $\varphi$ is a superposition
of positive-frequency solutions only.
Consequently, one can be certain that a particle will 
be detected at any fixed value of $t$ (assuming, of course, an ideal detector with perfect
efficiency).

Now we can generalize it to $n$-particle relativistic QM. 
Due to (\ref{consoverall}), the conservation equation (\ref{pr4.2}) generalizes to 
\begin{equation}\label{pr4.3}
 \frac{\partial |\varphi|^2}{\partial\tau} + 
\sum_a \partial_{a\mu} (|\varphi|^2 v^{\mu}_a)=0 ,
\end{equation}
where $v^{\mu}_a$ is the right-hand side of (\ref{bohm2n}).
Consequently, by analogy with (\ref{pr5.1}), (\ref{pr5.2}) generalizes to
\begin{equation}\label{pr5.3}
 p(x_1,\ldots,x_n)=|\varphi(x_1,\ldots,x_n)|^2 ,
\end{equation}
which is consistent with the Bohmian equation of motion (\ref{bohm2n}).
Similarly, by analogy with (\ref{pr8.1})-(\ref{pr9.1}), (\ref{pr8.2})-(\ref{pr9.2})
generalize to
\begin{equation}\label{pr8.3}
p_{t_1,\ldots,t_n}({\bf x}_1,\ldots,{\bf x}_n)
=\frac{|\varphi(t_1,{\bf x}_1,\ldots,t_n,{\bf x}_n)|^2}{N_{t_1,\ldots,t_n}} ,
\end{equation}
\begin{equation}\label{pr9.3}
N_{t_1,\ldots,t_n}=\int d^3x_1 \cdots \int d^3x_n
|\varphi(t_1,{\bf x}_1,\ldots,t_n,{\bf x}_n)|^2 .
\end{equation}
Eq. (\ref{pr8.3}) is nothing but the conditional probability density
associated with the {\it a priori} joint probability density (\ref{pr5.3}).
It can also be generalized to the case in which the hypersurfaces of
constant $x^0_a$ are replaced by arbitrary curved spacelike hypersurfaces
$\Sigma_a$. On such general hypersurfaces one needs to know 
the determinants of the induced metric $g^{(3)}_a({\bf x}_a)$, 
where ${\bf x}_a$ are the coordinates on $\Sigma_a$.
Eqs.~(\ref{pr8.3})-(\ref{pr9.3}) then further generalize to
\begin{equation}\label{pr8.4}
\tilde{p}_{\Sigma_1,\ldots,\Sigma_n}({\bf x}_1,\ldots,{\bf x}_n)
=\frac{\sqrt{|g^{(3)}_1({\bf x}_1)|} \cdots \sqrt{|g^{(3)}_n({\bf x}_n)|} \;
|\varphi(x_1,\ldots,x_n)|^2}{N_{\Sigma_1,\ldots,\Sigma_n}} ,
\end{equation}
\begin{equation}\label{pr9.4}
N_{\Sigma_1,\ldots,\Sigma_n}=\int_{\Sigma_1} d^3x_1 \sqrt{|g^{(3)}_1({\bf x}_1)|}
\cdots \int_{\Sigma_n} d^3x_n \sqrt{|g^{(3)}_n({\bf x}_n)|} \;
|\varphi(x_1,\ldots,x_n)|^2 .
\end{equation}
The tilde on $\tilde{p}$ denotes that it transforms as a scalar density, which in
curved coordinates should be distinguished from a quantity that transforms 
as a scalar.

What remains is to generalize all this to bosonic strings and superstrings.
The generalization to bosonic strings is trivial because all one needs to do is to
replace the discrete label $a$ by a continuous variable $\sigma$. 
In particular, the analog of (\ref{pr5.3}) is
\begin{equation}\label{pr5.4}
 p[X]=|\Psi[X]|^2 ,
\end{equation}
and similarly for the other equations. For superstrings, it further generalizes to
\begin{equation}\label{pr5.5}
 p[X]=\int[dM] \, |\Psi[M,X]|^2 .
\end{equation}
Indeed, $\tau$ in (\ref{stringbohm2.2}) can be rescaled, such that 
the Bohmian equation of motion becomes 
\begin{equation}\label{stringbohm2.3}
\frac{\partial X^{\mu}(\sigma,\tau)}{\partial\tau}=
\frac{ J^{\mu}[X;\sigma)}{p[X]} .
\end{equation}
This is consistent with the conservation equation
\begin{equation}\label{pr4.5}
 \frac{\partial p}{\partial\tau} + 
\int d\sigma \, \frac{\delta}{\delta X^{\mu}(\sigma)} J^{\mu}[X;\sigma)=0 ,
\end{equation}
that results from (\ref{consJ}).

Few additional remarks are in order. First, the probabilistic interpretation
(\ref{pr5.4}) requires that $\Psi[X]$ should be formally normalized such that
$\int [dX]\, |\Psi[X]|^2=1$. However, the integral $\int [dX]\, |\Psi[X]|^2$ is actually infinite, 
so such a normalization does not exist in a strict sense. Nevertheless, 
this integral can be made finite by replacing the measure $[dX]$ 
by an appropriately regularized measure. (For example, $\sigma$ can be replaced
by a discrete variable and the range of values of $X^{\mu}$ can be taken to be finite.)
At the end of calculation the regularization parameters can be turned back to their
natural values. One does not expect that such manipulations would
affect the final measurable results, but a detailed analysis of these technical 
subtleties is beyond the scope of the present paper.

Second, the probabilistic interpretation (\ref{pr5.4}) or (\ref{pr5.5}) is not the probabilistic
interpretation used in standard applications of string theory. 
In fact, in standard applications of string theory one does not discuss a probabilistic
interpretation in the configuration space at all.
Instead, one is usually 
interested in probabilities that the string will end up in a state $\Psi_A[X,M]$, where 
$\Psi_A[X,M]$ is a state that has definite values of particle-like properties 
(such as energy, momentum, and spin) interesting in
scattering experiments. By writing
\begin{equation}\label{meas1}
 \Psi[X,M]=\sum_A c_A \Psi_A[X,M] ,
\end{equation}
the appropriate probabilities are $p_A=|c_A|^2$. The standard applications of string theory
are concentrated on practical methods for calculation of the amplitudes $c_A$.
However, a complete quantum description
must also involve a description of the measuring apparatus. Thus, instead of
(\ref{meas1}), we actually have the total wave functional
\begin{equation}\label{meas2}
 \Psi_{\rm total}[X_{\rm total},M_{\rm total}]=\sum_A c_A \Psi_A[X,M]
\Phi_A[X_{\rm app},M_{\rm app}] ,
\end{equation} 
where $\Phi_A[X_{\rm app},M_{\rm app}]$ are normalized states of the measuring apparatus
that do not overlap in the $X_{\rm app}$-space. Consequently, the probabilistic
interpretation (\ref{pr5.5}) provides that the probability that $X_{\rm app}$ will 
take a value from the support of $\Phi_A[X_{\rm app},M_{\rm app}]$ is equal
to $|c_A|^2$. This shows that the probabilistic interpretation (\ref{pr5.4})
is compatible with the standard one.

Third, the probabilistic interpretation (\ref{pr5.5}) allows one additional argument
for the equation of motion (\ref{stringbohm2.3}). Even without the Bohmian interpretation,
the average local momentum of the string is
\begin{equation}\label{aver1}
 \langle P^{\mu}(\sigma) \rangle =
\int [dX] \int [dM] \, \Psi^*[X,M] \hat{P}^{\mu}(\sigma) \Psi[X,M] =
\int [dX] \frac{ J^{\mu}[X;\sigma)}{2} ,
\end{equation}
where (\ref{eqn34}) and (\ref{curbosGrS}) have been used to obtain the second equality.
We require that the average momentum of Bohmian trajectories should be equal to the
quantum-mechanical average momentum (\ref{aver1}). Since the Bohmian velocities
have the same directions in spacetime as Bohmian momenta, the simplest way
to accomplish this requirement is to postulate that the Bohmian velocities have the same
directions as $J^{\mu}$. This leads to (\ref{stringbohm2.2}) equivalent to
(\ref{stringbohm2.3}). Alternatively, instead of invoking the argument of 
simplicity, the same result can be obtained by applying a relativistic 
generalization of the argument in \cite{weak1,weak2} based on weak measurements of
velocities. Of course, analogue arguments can also be applied to bosonic string theory
without the $M$ variable.

Fourth, even though the Bohmian velocities (\ref{stringbohm2.3}) may be 
spacelike and past-oriented in some cases, no deviations from predictions 
of standard quantum theory are possible when these velocities are measured,
i.e., when the total state is of the form of (\ref{meas2}) with $\Psi_A$ being the
eigenstates of $\hat{P}^{\mu}$. In particular, superstring theory does not contain
tachyons, so measured velocities cannot be spacelike.

\section{Conclusion}
\label{SECCONCL}

This paper contains several new results on boson-fermion unification,
string theory, and Bohmian mechanics. As our main aim is to
emphasize the new conceptual aspects, some results can certainly be
formulated more rigorously in the future. 
But what exactly are the problems that these results are supposed to 
solve? It depends on what one considers to be a problem.

From the perspective of a traditional theoretician not interested
in foundations of quantum mechanics (QM), we have contributed to the 
understanding of boson-fermion unification by constructing
unifying particle and superstring currents. 
We have also 
%
seen that
these currents lead to a preferred 
spacetime foliation in effective field theory and indicated how it
reinterprets the notion of particles in QFT in curved spacetime.
Still, to fully understand the physical meaning of these currents, 
it is necessary to deal with foundations of QM itself.

From the perspective of physicists interested in various interpretations
of QM, we have presented a new argument for the naturalness of the 
Bohmian interpretation, by observing that this
is a natural interpretation of the unifying particle and superstring
currents. In addition, we have further developed a relativistic 
covariant probabilistic interpretation of wave functions based on
earlier ideas \cite{stuec,horw,kypr,fanchi}.

From the perspective of physicists interested in the internal problems
of the Bohmian interpretation, we have offered solutions to several
fundamental problems of that interpretation. These
problems and solutions may be summarized by the following items:
\begin{enumerate}
\item Fundamental ontology -- only strings (appearing as particles at 
low energies) have an objective existence, for both fermions 
and bosons.
\item Creation and destruction of particles --
corresponds to continuous string splitting described by the Bohmian
equation of motion.
\item Superluminal Bohmian velocities -- play an appealing role
for string splitting without singular splitting points.
\item Preferred foliation of spacetime --  
appears to be related (in a surprising way) to the boson-fermion unification.
\item Relativistic covariance -- once the particle or string wave function 
is known, the Bohmian equations of motion and the corresponding
probabilistic predictions are well defined and relativistic covariant.
\end{enumerate}  

Of course, string theory, supersymmetry, and boson-fermion unification
are still not experimentally confirmed, so other variants of the Bohmian
interpretation are not excluded by our theoretical results. Nevertheless,
from an aesthetical point of view, such a unified Bohmian picture seems to
be a very natural one. In addition,
as both superstring theory and Bohmian mechanics contain some 
nonlocal features that may manifest as a preferred foliation of spacetime,
it is possible that there is an even deeper not yet discovered relation
between these two theories.
Therefore, we believe that our results show
that physicists interested in the Bohmian interpretation should
become more interested in superstring theory. Conversely,
we also believe that the results of this paper, together with those
of \cite{nikstring1,nikdual}, suggest that physicists interested in
string theory should become more interested in Bohmian mechanics as well.
The fact that both theories are often unjustifiably
criticized for not being
experimentally testable is not the only thing that relates them.
Although the fundamental problem of measurement in QM is traditionally
not regarded as a problem that the ``theory of everything" is 
supposed to solve, our results on the relation between 
superstring theory and Bohmian mechanics suggest that the solution
of the measurement problem could also be a part of it.

\section*{Acknowledgement}
The author is grateful to the referees for their valuable suggestions and stimulating objections.
This work was supported by the Ministry of Science of the
Republic of Croatia under Contract No.~098-0982930-2864.

\end{document}